%% file: main_bib.tex
\documentclass[a4paper,10pt,conference]{IEEEtran}
\IEEEoverridecommandlockouts
\usepackage{cite}
\usepackage{amsmath,amssymb,amsfonts}
\usepackage{algorithmic}
\usepackage{url}
\usepackage{graphicx} 
\usepackage{color} 
\usepackage{textcomp}
\usepackage{xcolor}
\def\BibTeX{{\rm B\kern-.05em{\sc i\kern-.025em b}\kern-.08em
    T\kern-.1667em\lower.7ex\hbox{E}\kern-.125emX}}
\begin{document}

\title{Fake-image detection with Robust Hashing\\
{\footnotesize \textsuperscript{*}}
}

\author{\IEEEauthorblockN{1\textsuperscript{st} Miki Tanaka}
\IEEEauthorblockA{\textit{Tokyo Metropoliltan University} \\
Tokyo, Japan \\
tanaka-miki@ed.tmu.ac.jp}
\and
\IEEEauthorblockN{2\textsuperscript{nd} Hitoshi Kiya}
\IEEEauthorblockA{\textit{Tokyo Metropoliltan University} \\
Tokyo, Japan \\
kiya@tmu.ac.jp}
}

\maketitle

\begin{abstract}
In this paper, we investigate whether robust hashing has a possibility to robustly detect fake-images even when multiple manipulation techniques such as JPEG compression are applied to images for the first time. In an experiment, the proposed fake detection with robust hashing is demonstrated to outperform state-of-the-art one under the use of various datasets including fake images generated with GANs.
\end{abstract}


\section{Introduction}
Recent rapid advances in image manipulation tools and deep image synthesis techniques, such as Generative Adversarial Networks (GANs) have easily generated fake images. In addition, with the spread of SNS (social networking services), the existence of fake images has become a major threat to the credibility of the international community. Accordingly, detecting manipulated images has become an urgent issue\cite{outview}. 

Most forgery detection methods assume that images are generated by using a specific manipulation technique, and the methods aim to detect unique features caused by the manipulation technique such as checkerboard artifacts\cite{chekerboard1,checkerboard2,checkerboard3,chekerboard4}. 
Actually tampered images are usually uploaded to SNS and image sharing services. SNS providers are known to process the uploaded images by resizing or compressing them into JPEG format\cite{Socialmedia1,Socialmedia2,Socialmedia3,Socialmedia4}. Such manipulation may damage or lose the unique features of fake images.
However, the influence of manipulations on images has not been discussed sufficiently when a number of manipulation techniques such as JPEG compression are applied at the same time. 
In this paper, we investigate the possibility that there is a method with robust hashing that has been proposed for image retrieval, and the proposed method with robust hashing is demonstrated to have a high fake-detection accuracy, even when multiple manipulation techniques are carried out.

\section{Related work}
\subsection{Fake-image generation}
Fake images are manually generated by using image editing tools such as Photoshop. Splicing, copy-move, and deletion are also carried out under the use of such a tool. Similarly, resizing, rotating, blurring, and changing the color of an image can be manually carried out.

\begin{figure}[h]
    \centerline{\includegraphics[width=7cm]{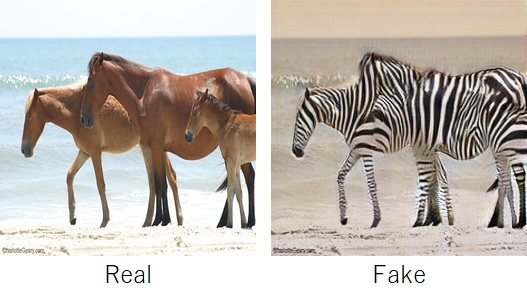}}
    \caption{Example Fake-images with CycleGAN}
    \label{fig}
\end{figure}

In addition, recent rapid advances in deep image synthesis techniques such as GANs have automatically generated fake images. CycleGAN\cite{cyclegan} and StarGAN\cite{stargan} are typical image synthesis techniques with GANs. CycleGAN is a GAN that performs one-to-one transformations, e.g. changing apples to oranges, while StarGAN is a GAN that performs many-to-many transformations, such as changing a person's facial expression or hair color (see Figs.\ref{fig} and \ref{fig:datasetsample}).
Furthermore, fake videos created using deep learning are called Deepfake, and various tampering methods have emerged, such as those using autoencoders, Face2Face\cite{deepfake_face2face}, FaceSwap\cite{deepfake_faceswapping}, and so on.

Real-world fake images may include the influence of a number of manipulation techniques such as image compression, resizing, copy-move at the same time, even if fake-images are generated by using GANs. Therefore, we have to consider such conditions for detecting real-world fake images.

\subsection{Fake detection methods}
Image tampering has a longer history than that of deep learning. Fragile watermarking\cite{fragile}, detection of double JPEG compression with a statistical method\cite{doubleJPEGcompression}\cite{doubleJPEGcompression2}, and use of PRNU (photo-response non-uniformity) patterns of each camera\cite{prnu1}\cite{prnu2} have been proposed to detect such tampers. However, most of them do not suppose to detect fake images generated with GANs. Moreover, they cannot detect the difference between fake images and just manipulated ones such as resized images, which are not fake images in general.

With the development of deep learning, fake detection methods with deep leaning have been studied so far. The methods with deep learning do not employ a reference image or the features of a reference image to detect tamper ones. The methods also assume that images are generated by using a specific manipulation technique to detect unique features caused by the manipulation technique.

There are several detection methods with deep learning for detecting fake images generated with an image editing tool as Photoshop. Some of them focus on detecting the boundary between tampered regions and an original image\cite{highpath}\cite{LSTM}\cite{2stream}. Besides, a detection  method\cite{exif} enables us to train a model without tamper images.

Most detection methods with deep learning have been proposed to detect fake images generated by using GANs. An image classifier trained only with ProGAN was shown to be effective in detecting images generated by other GAN models\cite{CNN-gimg}.
Various studies have focused on detecting checkerboard artifacts caused in both of two processes: forward propagation of upsampling layers and backpropagation of convolutional layers\cite{AutoGAN}. In this work, the spectrum of images is used as an input image in order to capture the checkerboard artifacts. 

To detect fake videos called DeepFake, a number of detection methods have been investigated so far. Some methods attempt to detect failures in the generation of fake videos, in terms of poorly generated eyes and teeth\cite{deepfake_visualartifact}, the frequency of blinking as a feature\cite{deepfake_blink}, and the correctness of facial landmarks\cite{deepfake_landmark} or head posture\cite{deepfake_headpose}. However, all of these methods have been pointed out to have problems in the robustness against the difference between training datasets and test data\cite{outview}. In addition, the conventional methods have not considered the robustness against the combination of various manipulations such as the combination of resizing and DeepFake.

\section{Proposed method with Robust Hashing}
\subsection{Overview}
\begin{figure}[h]
    \centerline{\includegraphics[width=8cm]{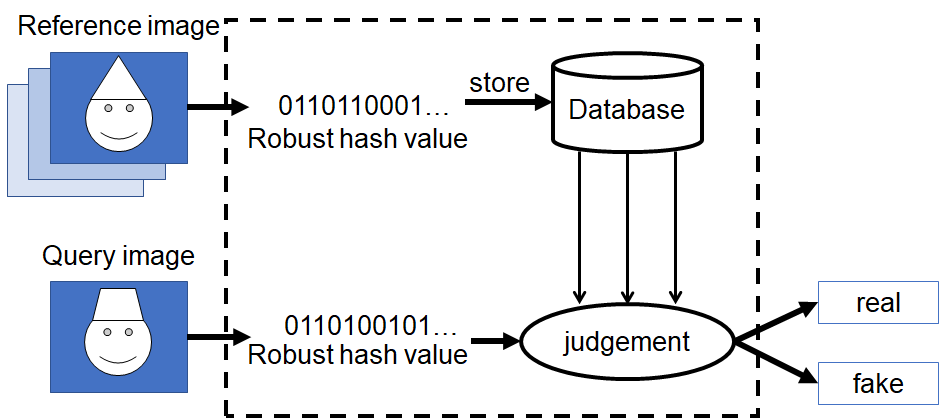}}
    \caption{Overview of proposed method}
    \label{fig:overview}
\end{figure}

Figure\ref{fig:overview} shows an overview of the proposed method. In the framework, robust hash value is computed from easy reference image by using a robust hash method, and stored in a database. Similar to reference images, a robust hash value is computed from a query one by using the same hash method. The hash value of the query is compared with those stored the database. Finally, the query image is judged whether it is real or fake in accordance with the distance between two hash values.

\subsection{Fake detection with Robust Hashing}
Various robust hashing methods have been proposed to retrieval similar images to a query one\cite{hashQuaternion, robusthash}.
In this paper, we apply the robust hashing method proposed by Li et al\cite{hashQuaternion} for applying it to fake-image detection. This robust hashing enables us to robustly retrieve images, and has the following properties.
\begin{itemize}
    \item Resizing images to 128$\times$128 pixels prior to feature extraction.
    \item Performing 5$\times$5-Gaussian low-pass filtering with a standard deviation of 1.
    \item Using rich features extracted from spatial and chromatic characteristics.
    \item Outputting a bit string with a length of 120 bits as a hash value.
\end{itemize}
In the method, the similarity is evaluated in accordance with the hamming distance between the hash string of a query image and that of each image in a database. 

Let vectors $\boldsymbol{u}=\{u_1,u_2,\dots,u_n\}$ and $\boldsymbol{q}=\{q_1,q_2,\dots,q_n\}$, $u_i,\ q_i \in \{0,1\}$ be the hash strings of reference image $U$ and query image $Q$, respectively. The hamming distance $d_H(\boldsymbol{u},\boldsymbol{q})$ between $U$ and $Q$ is given by:
\begin{eqnarray}
    \label{eq_1}
    d_H(\boldsymbol{u},\boldsymbol{q}) \triangleq \sum^n_{i=1}\delta(u_i,q_i)
\end{eqnarray}
where
\begin{eqnarray}
    \label{eq_2}
    \delta(u_i,q_i)=\left\{ \begin{array}{ll}
        0,\ u_i=q_i \\
        1,\ u_i\neq q_i \\
    \end{array} \right. .
\end{eqnarray}
To apply this similarity to fake-image detection, we introduce a threshold $d$ as follows.
\begin{eqnarray}
    \label{eq_3}
    \begin{cases}
        Q \in \mathbb U', \underset{u\neq q, u\in \mathbb U}{\min}(d_H(\boldsymbol{u},\boldsymbol{q})) < d \\
        Q \notin \mathbb{U}', \underset{u\neq q, u\in \mathbb{U}}{\min}(d_H(\boldsymbol{u},\boldsymbol{q})) \geq d
    \end{cases}
\end{eqnarray}
where $\mathbb U$ is a set of reference images and $\mathbb U'$ is the an of images generated with image manipulations from $\mathbb U$, which does not include fake images. According to eq. (\ref{eq_3}), $Q$ is judged whether it is a fake image or not.

\section{Experiment results}
The proposed fake-image detection with robust hashing was experimentally evaluated in terms of accuracy and robustness against image manipulations.
\subsection{Experiment setup}
\begin{figure}[h]
    \centerline{\includegraphics[width=7cm]{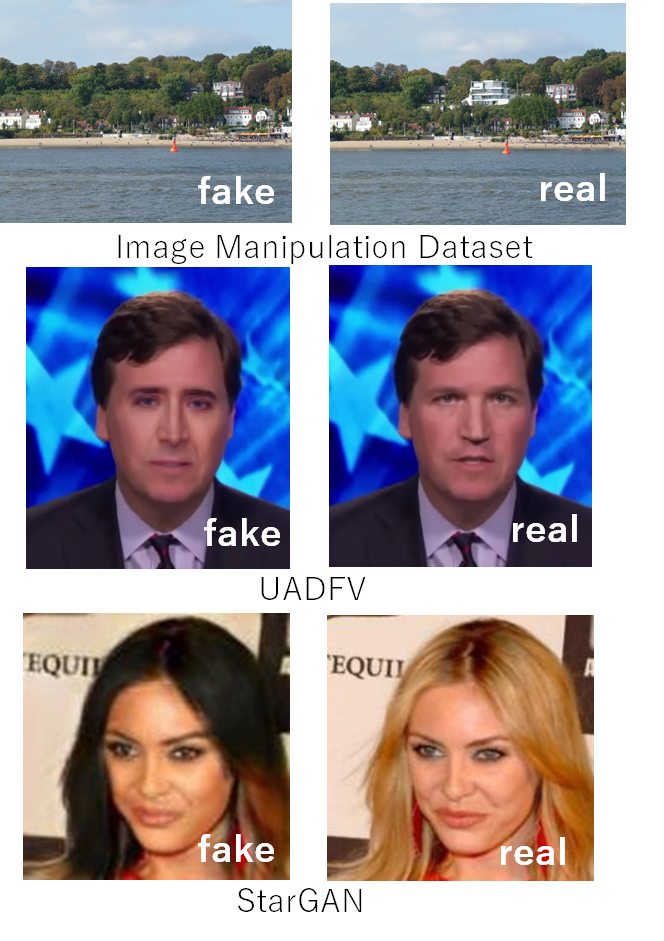}}
    \caption{Example of datasets}
    \label{fig:datasetsample}
\end{figure}

In the experiment, four fake-image datasets: Image Manipulation Dataset\cite{image_m_d}, UADFV\cite{deepfake_blink}, CycleGAN\cite{cyclegan}, and StarGAN\cite{stargan} were used. The details of datasets are shown in Table \ref{tb_dataset} (see Figs. \ref{fig} and \ref{fig:datasetsample}). The datasets consist of pairs of a fake-image and the original one. JPEG compression with a quantization parameter of $Q_J=80$ was applied to all query images. $d = 3$ was selected as threshold $d$ in accordance with the EER (Equal error rate) performance.
\begin{table}[h]
    \caption{Datasets}
    \label{tb_dataset}
    \begin{center}
        \begin{tabular}{c||c|c|c}
        \hline
        dataset & Fake-image generation& real
        & fake\\ \cline{3-4}
        & \ & \multicolumn{2}{|c}{No. of images}  \\ \hline
        \hline
            \begin{tabular}{c}
                Image \\
                Manipulation \\
                Dataset\cite{image_m_d} 
            \end{tabular} & copy-move & 48 & 48\\ \hline
            UADFV\cite{deepfake_blink} & face swap & 49 & 49\\ \hline
            CycleGAN\cite{cyclegan} & GAN & 1320 & 1320\\ \hline
            StarGAN\cite{stargan} & GAN & 1999 & 1999\\ 
        \hline
        \end{tabular}%
    \end{center}
\end{table}

As one of the state-of-the-art fake detection methods, Wang's method\cite{CNN-gimg} was compared with the proposed one. Wang's method was proposed for detecting images generated by using CNNs including various GAN models, where a classifier is trained by using ProGAN. 

The performance of fake-image detection was evaluated by using AP (Average Precision) and Accuracy (fake), given by,
\begin{eqnarray}
    \label{eq:accuracysample}
    Accuracy\ (fake) = \frac{N_{tn}}{N_{Qf}}
\end{eqnarray}
where $N_{Qf}$ is the number of fake query images, and $N_{tn}$  is the number of fake query ones that are correctly judged as fake images.

\subsection{Results without additional manipulation}
Table \ref{tb_gan} shows experimental results under the use of the two detection methods. 
From the table, it is shown that the proposed method had a higher performance than Wang's method in terms of both AP and Acc (fake). In addition, the performance of Wang's method heavily decreased when using the image manipulation and UADFV datasets. The reason is that Wang's method focuses on detecting fake images generated by using CNNs. The image manipulation dataset does not consist of images generated with GANs. In addition, although UADFV consists of images generated by using DeepFake, they have the influence of video compression.

\begin{table}[h]
    \caption{comparison with Wang's method}
    \label{tb_gan}
    \begin{center}
        \scalebox{0.90}{
    \begin{tabular}{c||c|c|c|c}
    \hline
    & \multicolumn{2}{|c}{Wang's method\cite{CNN-gimg}}& \multicolumn{2}{|c}{proposed}  \\ \cline{2-5}
    Dataset & AP & Acc\ (fake) & AP & Acc\ (fake) \\
    \hline
    Image Manipulation Dataset & 0.5185 & 0.0000 & 0.9760 & 0.8750\\
    UADFV & 0.5707 & 0.0000 & 0.8801 & 0.7083\\
    CycleGAN & 0.9768 & 0.5939 & 1.0000 & 1.0000\\
    StarGAN& 0.9594 & 0.5918 & 1.0000 & 1.0000\\
    \hline
    \end{tabular}
        }
    \end{center}
\end{table}

\subsection{Results with additional manipulation}
JPEG compression with $Q_J=70$, resizing with a scale factor of 0.5, copy-move or splicing was applied to query images. Therefore, when query images were fake ones, the fake query ones included the effects of two manipulations at the same time.

Table \ref{tb_gancopymove} shows experimental results under the additional manipulation, where 50 fake images generated by using CycleGAN, in which horses were converted to zebras, were used (see Fig.\ref{fig}). The proposed method was confirmed to still maintain a high accuracy even under the additional manipulation. In contrast, Wang's method suffered from the influence of the addition manipulation. In particular, for splicing and resizing, Wang's method was affected by these operations. That is why the method assume that fake images are generated by using CNNs, to detect unique features caused by using CNNs. However, splicing and resizing don't depend on CNNs, although CycleGAN includes CNNs.

\begin{table}[h]
    \caption{Comparison with Wang's method under additional manipulation (dataset: CycleGAN)}
    \label{tb_gancopymove}
    \begin{center}
        \scalebox{0.90}{
    \begin{tabular}{c||c|c|c|c}
    \hline
    & \multicolumn{2}{|c}{Wang's method\cite{CNN-gimg}}& \multicolumn{2}{|c}{proposed}  \\ \cline{2-5}
    additional manipulation & AP & Acc\ (fake) & AP & Acc\ (fake) \\ \hline
    None & 0.9833 & 0.6200 & 0.9941 & 1.0000\\
    JPEG($Q_J=70$) & 0.9670 & 0.6000  & 0.9922 & 0.9800\\
    resize\ (0.5) & 0.8264 & 0.2400 & 0.9793 & 1.0000\\
    copy-move& 0.9781 & 0.6000 & 1.0000 & 1.0000\\
    splicing& 0.9666 & 0.4800 & 0.9992 & 1.0000\\
    \hline
    \end{tabular}%
        }
    \end{center}
\end{table}

\section{Conclusion}
In this paper, we proposed a novel fake-image detection method with robust hashing for the first time. Although various robust hashing methods have been proposed to retrieve similar images to a query one so far, a robust hashing method proposed by Li et al was applied to various datasets including fake images generated with GANs. In the experiment, the proposed method was demonstrated not only to outperform a state-of-the-art but also to be robust against the combination of image manipulations.

\input{main_bib.bbl}


\end{document}